 \documentclass[preprint2]{aastex}
\newcommand{\etal}{{\em et~al.}}
\newcommand{\ie}{{\em i.e.}}
\newcommand{\RSun}{R_{\odot}}
\newcommand{\rSun}{r/\RSun}
\newcommand{\Aone}  {${\sf A}_1$}
\newcommand{\Atwo}  {${\sf A}_2$}
\newcommand{\Athree}{${\sf A}_3$}
\newcommand{\Afour} {${\sf A}_4$}

\begin{document}

\title{Sensitivity analysis of the solar rotation to helioseismic data from
GONG, GOLF and MDI observations.}

\author{A.~Eff-Darwich\altaffilmark{1,2}}

\affil{$^1$Dept.~Edafolog\'\i a y Geolog\'\i a, Universidad de La Laguna,
Tenerife, 38205, Spain}

\affil{$^2$Instituto de Astrof\'\i sica de Canarias, C/ V\'\i a L\'actea s/n,
Tenerife, 38205, Spain}

\author{S.~G.~Korzennik\altaffilmark{3}}

\affil{$^3$Harvard-Smithsonian Center for Astrophysics, 60 Garden St.,
Cambridge, MA, 02138}

\author{S.~J.~Jim\'enez-Reyes\altaffilmark{2,4}}
\affil{$^2$Instituto de Astrof\'\i sica de Canarias, C/ V\'\i a L\'actea s/n, 
Tenerife, 38205, Spain}
\affil{$^4$School of Physics and Astronomy, University of Birmingham,
Edgbaston, Birmingham B15 2TT,UK.}
\author{R.A. Garc\'\i a \altaffilmark{5,6}}

\affil{$^5$Service d'Astrophysique CEA/DSM/DAPNIA, CE Saclay, 
91191 Gif-sur-Yvette CEDEX, France}

\affil{$^6$AIM - Unit\'e Mixte de Recherche CEA - CNRS - Universit\'e Paris VII -
 UMR n$^o$ 7158, CEA Saclay, 
91191 Gif-sur-Yvette CEDEX, France}

\begin{abstract}
Accurate determination of the rotation rate in the radiative zone of the sun
from helioseismic observations requires rotational frequency splittings of
exceptional quality as well as reliable inversion techniques. We present here
inferences based on mode parameters calculated from 2088-days long MDI, GONG
and GOLF time series that were fitted to estimate very low frequency rotational
splittings ($\nu < 1.7$ mHz). These low frequency modes provide data of
exceptional quality, since the width of the mode peaks is much smaller than
the rotational splitting and hence it is much easier to separate the
rotational splittings from the effects caused by the finite lifetime and the
stochastic excitation of the modes.

We also have implemented a new inversion methodology that allows us to infer
the rotation rate of the radiative interior from mode sets that span $\ell=1$
to 25. Our results are compatible with the sun rotating like a rigid solid in
most of the radiative zone and slowing down in the core ($\rSun < 0.2$). A
resolution analysis of the inversion was carried out for the solar rotation
inverse problem.  This analysis effectively establishes a direct relationship
between the mode set included in the inversion and the sensitivity and
information content of the resulting inferences.  We show that such an
approach allows us to determine the effect of adding low frequency and low
degree $p$-modes, high frequency and low degree $p$-modes, as well as some
$g$-modes on the derived rotation rate in the solar radiative zone, and in
particular the solar core. We conclude that the level of uncertainties that is
needed to infer the dynamical conditions in the core when only $p$-modes are
included is unlikely to be reached in the near future, and hence sustained
efforts are needed towards the detection and characterization of $g$-modes.
\end{abstract}

\section{Introduction}

Over the past decade, increasingly accurate helioseismic observations from
ground-based and space-based instruments have given us a reasonably good
description of the dynamics of the solar interior \citep[e.g.][]{bi50,
thompson2003}.  Helioseismic inferences have confirmed that the differential
rotation observed at the surface persists throughout the convection
zone. There appears to be very little, if any, variation of the rotation rate
with latitude in the outer radiative zone ($0.4 > \rSun > 0.7$). In that
region the rotation rate is almost constant ($\approx 430$ nHz), while at the
base of the convection zone, a shear layer ---known as the tachocline---
separates the region of differential rotation throughout the convection zone
from the one with rigid rotation in the radiative zone.

 Despite the large scatter amongst the rotational splittings that are
sensitive to the solar core \citep[see discussion in][]{Eff-Darwich2002}, we
can rule out an inward increase or decrease of the solar internal rotation
rate down to $\rSun \approx 0.25$, by more than $20 \%$ of the surface rate at
mid-latitude \citep{bi3, Eff-Darwich2002, couvidat2003}.  This is in clear
disagreement with the theoretical hydrodynamical models that predict a much
faster rotation in the solar core, namely 10 to 50 times faster than the
surface rate \citep[e.g.][]{thompson2003}.
 
  More recently, \cite{garcia2004} and \cite{korzennik2005} have independently
developed new mode fitting procedures to improve the quality and precision of
the characterization of the modes that are sensitive to the rotation in the
solar core. By using very long time series ---spanning nearly six years of
observations--- collected with the MDI \citep{mdi}, GONG \citep{gong} and GOLF
\citep{golf} instruments they have measured rotational splittings for modes
with frequencies as low as $1.1$ mHz.

  We present here an attempt to constraint the radial and latitudinal
distribution of the rotation rate in the radiative interior through the
inversion of a combined MDI, GONG \& GOLF data set. We also attempt to
establish the sensitivity of helioseismic data sets to the dynamics of the
inner solar radiative interior, as well as the level of accuracy that
helioseismic data should have to resolve the solar core.

\section{Theoretical background}

  The starting point of all linear rotational helioseismic inversion
methodologies is the functional form of the perturbation in frequency, $\Delta
\nu _{n \ell m}$, induced by the rotation of the sun, $\Omega(r,\theta)$, and
given by \citep[see derivation in][]{bi46}:
\begin{equation} 
\Delta \nu _{n \ell m} = \frac{1}{2\pi}\int_0^R \int_0^{\pi}
K_{n \ell m}(r,\theta)\Omega(r,\theta)drd\theta \pm \epsilon_{n \ell m}
\label{eq:equation4} 
\end{equation}

  The perturbation in frequency, $\Delta \nu _{n \ell m}$, with the
observational error, $\epsilon_{n \ell m}$, that corresponds to the rotational
component of the frequency splittings, is given by the integral of the product
of a sensitivity function, or kernel, $K_{n \ell m}(r,\theta)$, with the
rotation rate, $\Omega(r,\theta)$, over the radius, $r$, and the co-latitude,
$\theta$. The kernels, $K_{n \ell m}(r,\theta)$, are known functions of the
solar model.

  Equation \ref{eq:equation4} defines a classical inverse problem for the
sun's rotation. The inversion of this set of $M$ integral equations -- one for
each measured $\Delta \nu _{n \ell m}$ -- allows us to infer the rotation rate
profile as a function of radius and latitude from a set of observed rotational
frequency splittings (hereafter referred as splittings).

  The inversion method we use is based on the Regularized Least-Squares
methodology (RLS). The RLS method requires the discretization of the integral
relation to be inverted. In our case, Eq.~\ref{eq:equation4} is transformed
into a matrix relation
\begin{equation}
  D = A x + \epsilon \label{eq:equation5}
\end{equation}
where $D$ is the data vector, with elements $\Delta \nu _{n \ell m}$ and
dimension $M$, $x$ is the solution vector to be determined at $N$ model grid
points, $A$ is the matrix with the kernels, of dimension $M \times N$ and
$\epsilon$ is the vector containing the corresponding observational
uncertainties.

   The RLS solution is the one that minimizes the quadratic difference
$\chi^2=|Ax-D|^2$, with a constraint given by a smoothing matrix, $H=G^TG$,
introduced in order to lift the singular nature of the problem \citep[see, for
additional details,][]{Eff-Darwich1997}. The matrix $G$ represents the
first-order discrete differential operator, although it will be shown below
that the inversion technique we have developed is to a first order
approximation independent of the choice of $G$. The general relation to be
minimized is
\begin{equation}
  S(x) = (Ax-D)^T(Ax-D) + \gamma xHx
\end{equation}
where $\gamma$ is a scalar introduced to give a suitable weight to the
constraint matrix $H$ on the solution. Hence, the function $x$ is approximated
by 
\begin{equation}
  x_{\rm est} = (A^TA + \gamma H)^{-1}A^TD  \label{eq:equation6}
\end{equation}

Replacing $D$ from Eq.~\ref{eq:equation5} we obtain

\begin{equation}
  x_{\rm est} = (A^TA + \gamma H)^{-1}A^TAx \stackrel{\mathrm{def}}{=} Rx  
\label{eq:equation7}
\end{equation}
hence
\begin{equation}
R = (A^TA + \gamma H)^{-1}A^TA \label{eq:equation8}
\end{equation}

The matrix $R$, that combines forward and inverse mapping, is referred to as
the resolution or sensitivity matrix \citep{friedel2003}. Ideally, $R$ would be
the identity matrix, which corresponds to perfect resolution. However, if we
try to find an inverse with a resolution matrix $R$ close to the identity, the
solution is generally dominated by the noise magnification. The individual
columns of $R$ display how anomalies in the corresponding model are imaged by
the combined effect of measurement and inversion. In this sense, each element
$R_{ij}$ reveals how much of the anomaly in the $j^{th}$ inversion model grid
point is transferred into the $i^{th}$ grid point. Consequently, the diagonal
elements $R_{ii}$ states how much of the information is saved in the model
estimate and may be interpreted as the resolvability or sensitivity of
$x_{i}$. We defined the sensitivity $\lambda_{i}$ of the grid point $x_{i}$
to the inversion process as follows:

\begin{equation}
\lambda_i = \frac{R_{ii}}{\sum_{j=1}^{N}R_{ij}} \label{eq:equation10}
\end{equation}

With this definition, a lower value of $\lambda_i$ means a lower sensitivity
of $x_{i}$ to the inversion of the solar rotation. We define a smoothing
vector $W$ with elements $w_i=\lambda_i^{-1}$ that is introduced in
Eq.~\ref{eq:equation6} to complement the smoothing parameter $\gamma$, namely

\begin{equation}
  x_{\rm est} = (A^TA + \gamma W H)^{-1}A^TD  \label{eq:equation11}
\end{equation}

Such substitution allows to apply different regularizations to different model
grid points $x_{i}$ whose sensitivities depend on the data set that are used
in the inversions. In this sense, the inversion is a two steps process: first
$R$ is obtained from Eq.~\ref{eq:equation7} for a small value of the
regularization parameter $\gamma$.  Then, the smoothing vector $W$ is
calculated through Eq.~\ref{eq:equation10} and the inversion estimates are
obtained through Eq.~\ref{eq:equation11}. A set of results can be calculated
for different values of $\gamma$, the optimal solution being the one with the
best trade-off between error propagation and the quadratic difference
$\chi^2=|Ax_{\rm est}-D|^2$ as introduced in \cite{Eff-Darwich1997}.

  In this paper we show how to use the matrix $R$ to study the sensitivity of
helioseismic data sets to the rotation rate of the solar interior.
Consequently, we present a theoretical analysis of the effect of adding low
frequency and low degree $p$-modes, high frequency and low degree $p$-modes,
and $g$-modes on the rotation rate of the solar core derived through numerical
helioseismic inversion techniques.

\section{Observational mode parameters and inversion results from 2088-days
long MDI, GOLF and GONG time series}

The work presented here is based on rotational frequency splittings measured
from observations by the GONG ground-based network and the MDI and GOLF
experiments on board the SOHO spacecraft.  All rotational splittings were
computed from 2088-days long time series, starting April 30th 1996 and ending
January 17th 2002, as summarized in Table~\ref{tab:table1}.

The three data sets {\sf KM, KG \& GG} (see Table~\ref{tab:table1} for
explanations) contain for the first time very low frequency rotational
splittings ($\nu < 1.7$ mHz). These low frequency modes provide data of
exceptional quality, since the width of the mode peaks is much smaller than
the rotational splitting. It is therefore much easier to separate the
rotational splittings from the effects caused by the finite lifetime and the
stochastic excitation of the modes. The data set {\sf SM} (see again
Table~\ref{tab:table1}) was obtained by averaging all the data sets resulting
from fitting the 72-days long MDI time series \citep{bi50} that overlap the
April 30th 1996 to January 17th 2002 period. The averaging process reduces
significantly the number of $\ell < 8$ modes in that data set.
 
  Since these data sets have been calculated from different time-series and
peak-fitting techniques, one can expect some differences among the different
data sets. When using different time-series, the mode parameters can be
affected by the changing solar activity cycle. Moreover, fitting techniques
can give different results if they are applied to either individual peaks or
to ridges \citep{korzennik2005}.  Differences between MDI, GOLF and GONG can
also arise from systematics introduced by the merging process used by GONG to
obtain single time-series from multiple stations located
worldwide. Differences may also come from the different spatial filters and
leakage matrices that are used to isolate the signal of an individual mode
\citep{korzennik2005, chaplin2006}. In any case, we combined the various data
sets in a single set -- following the prescription described in
Table~\ref{tab:table2}. Our newly developed inversion methodology was
applied to the combined set to infer the rotation rate in the solar interior.

  Figure~1 shows the observational frequency splitting uncertainties of the
combined data set as a function of radial order and degree, whereas Figs.~2
and 3 show the splittings uncertainties for sectoral modes as a function of a
proxy of the inner turning point of the modes, $\ell/\nu$, and as a function
of frequency, $\nu$, respectively. These plots clearly illustrate the well
known and challenging fact that only a small number of modes penetrate the
solar core and that the largest uncertainties are associated with these
modes. Indeed, the combined data set does not include low degree high
frequency modes (\ie, $\ell < 4$ and $\nu > 2.2$ mHz), since at higher
frequencies unwanted bias appears in the estimated splittings due to the
difficulty in separating the effect of rotational splitting from the limited
lifetime of the modes \citep{appour2000,chaplin2006}.

   The inversion of the combined MDI-GOLF-GONG data (see Fig.~4) confirms that
the well-known differential rotation observed at the surface persists
throughout the convection zone. Although only $\ell < 25$ modes were used in
the inversion, it was possible to infer the rotation rate in the convection
zone as the result of the exceptional quality of the low frequency splittings
($\nu < 1.7$ mHz) obtained by \citet{korzennik2005}. The differential rotation
changes abruptly at approximately $0.7 \RSun$ to rigid rotation throughout
the radiative zone. The radial distribution of the rotation is approximately
flat, at a rate of $\approx 430$ nHz, decreasing below $0.2 \RSun$. The
tendency below $0.15 \RSun$ is not real and results from extrapolation of the
trend seen at larger radii, as explained in the following section.

\section{Sensitivity analysis for the inversion of the solar radiative
interior.}

A theoretical analysis was carried out in order to determine the effect of
different low degree mode sets on the derivations of the solar rotation rate
of the inner radiative interior. Four different artificial data sets,
hereafter referred as {\Aone} to {\Afour} were calculated using
Eq.~\ref{eq:equation4} and an artificial rotation rate $\Omega_{\rm
A}(r,\theta)$ that is shown in Fig.~5.  The different artificial data sets
correspond to different mode sets and/or uncertainties, as explained in
Table~3. The observational uncertainties (standard errors) were taken from the
combined mode set used in the previous section, whereas the noise added to the
artificial data was calculated from normal distribution with the observed
uncertainties.  The {\Aone} data set contains the same mode set than the
combined MDI-GOLF-GONG data set.  Errors for g modes were arbitrarily set to 6
nHz (the mean of the observational uncertainties for the acoustic mode
splittings), since at present there are not reliable estimates for the
uncertainties of g-modes frequency splittings. In any case, we are interested
in the behavior of the inversion methodology when g modes are added, rather
than in the results of the inversion for different values of the splittings
and the observational errors.

The sensitivity vector, $\Lambda$, was computed for the four artificial data
sets, as illustrated in Figs.~6 and 7, where the sensitivities,
${\lambda_i}$, for the rotation rate in the solar interior are
presented as a function of the radius, for each artificial data set at the
equator or for several latitudes for a given set.  The data sets {\Aone} and
{\Atwo} are significantly less sensitive to the rotation of the solar core
than the other sets. Although the {\Atwo} set includes the same high frequency
modes as set {\Athree}, the errors of the {\Atwo} set
are significantly larger and hence the sensitivities do not differ from those
obtained for the {\Aone} set. The addition of two $g$-modes (in set {\Afour})
increases significantly the sensitivity to the solar core. However, it is
important to notice that even with the addition of $g$-modes, the sensitivity
at the solar core varies with the latitude, as illustrated in Fig.~7. For all
sets, the sensitivities to the equatorial regions of the solar core are larger
than the sensitivities at other latitudes.

The effect on the sensitivity vector of the choice of the smoothing matrix $H$
is presented in Fig.~8, where the equatorial sensitivities ${\lambda_i}$ for
the first, second and third-order discrete differential operators $G$ are
shown. The larger the order of the operator $G$ the larger the sensitivity,
except near the edges, and in particular the core. The smoothing vector $W$ is
obtained from the sensitivity vector $\Lambda$ and hence, the regularization
constraint will not only depend on the mode set, but also on the shape of $G$.

The choice of the number and spatial distribution of the model grid points,
$N$, is an important aspect of the inversion process. In non-adaptive
regularization inversions, decreasing the number of grid points is in itself a
form of regularization. The inversion procedure described here will also
adjust to the distribution of grid points. This is illustrated in Fig.~9,
where we show that the variation of the inversion sensitivities (and hence the
regularization constraints) is not constant with radius and latitude when the
number of grid points are changed. In this sense, the {\it a priori} choice of
the distribution of model grid points will not constrain the inversion
results.

The inversion methodology developed for the work presented here differs from
standard RLS techniques by introducing a smoothing vector $W$. The purpose of
this vector is to avoid over-smoothing the inversion solution in certain
regions of the solar interior and hence loose valuable information. This is
illustrated in Fig.~10, where two different inversions of the same data set
are presented, namely a standard RLS inversion (no vector $W$ is added) dotted
line, and the newly developed RLS inversion, solid line. The standard RLS
inversion tends to over-smooth and hence to assign unrealistic low errors to
the estimated rotation in the convection zone to get a stable solution in the
radiative regions. However when $W$ is added to the inversion procedure, the
over-smoothing problem in the convection zone is mitigated, since the
sensitivities ${\lambda_i}$ in the convection zone are larger and hence, the
smoothing coefficients ${w_i}$ are lower. The standard RLS technique would
assign the same smoothing coefficients to all model grid points in the
inversion.

The correspondence between the sensitivity analysis shown in Figs.~6 and 7 and
the information contained in the resolution matrix $R$ is illustrated in
Figs.~11 to 13, where we show the resolution vector corresponding to the
estimate at the equator for $r_i=0.06 \RSun$. This corresponds to the
averaging kernel defined by \cite{backus1970} for the Optimal Localized
Averages technique. In the ideal case, the resolution should be unity at the
location where the solution is estimated and zero elsewhere. Only in the
inversion of the {\Afour} set is the largest amplitude of the resolution
vector centered at $r_i=0.06 \RSun$, and thus the result is reliable. The poor
localization of the resolution at $r_i=0.06 \RSun$ for the inversions of the
{\Aone} to {\Athree} set could not be improved by any inversion
technique. However, this lack of resolution is taken into account by the
sensitivity analysis, since it evaluates low sensitivities and hence assigns
large regularizations to those grid points.
At larger radii both the location and the amplitude of the resolution are
significantly increased, as illustrated in Fig.~14, where we show the
resolution corresponding to the estimate at the equator for $r_i=0.2 \RSun$
obtained with the data set {\Athree}.

The conclusions derived from Figs.~6 and 7 can also be drawn from the
inversions of the data sets, as illustrated in Figs.~15 to 18.  Figures~15 and
16 show the inverted profile and error distribution for the {\Aone} set at
several latitudes and demonstrates that there is good sensitivity to the
latitudinal variation in the radiative rotation rate above $r \approx 0.3
\RSun$. Hence, the absence of differential rotation for the solar rotation
rate in the radiative interior (Fig.~4) is real, not an artifact of the
inversion procedure. The unrealistic flat rotation rate below $r \approx 0.15
\RSun$ resulting from the inversions of the {\Aone} and {\Atwo} sets is due to
the lack of sensitivity of the mode set to that region. As a result the
inversion extrapolates the trend of the solution at larger radii. There are
not significant differences in the inversions of sets {\Aone} and {\Atwo},
although set {\Atwo} includes low degree and high frequency modes. However,
the new information contained in the low degree and high frequency modes of
set {\Atwo} is lost due to they large observational uncertainties.  In all
four cases, larger differences between the artificial and the inverted
rotation rates are seen at higher latitudes, especially in the radiative
interior, as a result of the lack of sensitivity of the inversions to the
polar regions.

Only in the cases of sets {\Athree} and {\Afour} (see Fig.~17), was it
possible to infer the main trends of the rotation rate below $r \approx 0.15
\RSun$. However, it was necessary to include, either, data with unrealistic
small observational errors (set {\Athree}), or, a couple of $g$-modes (set
{\Afour}), modes that have yet to be unambiguously observed.  The most likely
way to reduce the observational uncertainties consists of increasing the
length of the time series. Figure~19 compares the observational errors for the
$\ell=25$ sectoral modes for five 728-days long data sets to the 2088-days
long data set, all estimated by \cite{korzennik2005}. The formal observational
uncertainties are proportional to the square root of the length of the time
series, hence it would be necessary to observe for decades to reduce the
observational uncertainties of the very low degree modes to the current levels
of the $\ell=25$ modes, all the while assuming that we can also reduce the
residual bias in our current estimates of the low degree and high frequency
splittings \citep[see discussions in][]{appour2000,chaplin2006}.

\section{Conclusions}

We have used for the first time a combined MDI-GOLF-GONG data set of
rotational frequency splittings that covers the largest possible frequency
range, spanning from 1.1 mHz to 3.9 mHz. This mode set was determined from
2088-days long time series acquired by the MDI, GOLF and GONG instruments and
analyzed independently by several authors, namely \citet{korzennik2005},
\citet{garcia2004}, \citet{gelly2002}, \citet{bi50} and \citet{jim}. Very low
frequency splittings ($\nu < 1.7$ mHz) were included to improve the precision
and the resolution of the inversion in the solar interior.

In order to optimally invert this unique data set, we implemented a new
inversion methodology that combines the regularized least-squares technique
with the analysis of the sensitivity of the solution at all model grid point
to the mode set being inverted.  The inversion of the actual MDI-GOLF-GONG
data set reveals that the sun rotates as a rigid solid in most of the
radiative interior and slowing down below $0.2 \RSun$.

  The calculation of the sensitivity vector $\Lambda$ offers a rapid and
intuitive way of evaluating the sensitivity of helioseismic data to the
dynamics of the solar interior, in particular in the core ($r < 0.25
\RSun$). We conclude that with the present accuracy of the available
splittings, it is not possible to derive the dynamical conditions below $r
\approx 0.2 \RSun$. This results from the relatively large observational
uncertainties of the modes sensitive to the solar core, in particular the low
degree and high frequency modes. The level of uncertainties that is needed to
infer the dynamical conditions in the core when only including $p$-modes is
unlikely to be reached in the near future, and hence sustained efforts are
needed towards the detection and characterization of $g$ modes.

\section{Acknowledgments}

  The Solar Oscillations Investigation - Michelson Doppler Imager project on
SOHO is supported by NASA grant NAS5--3077 at Stanford University.  SOHO is a
project of international cooperation between ESA and NASA.

  The GONG project is funded by the National Science Foundation through the
National Solar Observatory, a division of the National Optical Astronomy
Observatories, which is operated under a cooperative agreement between the
Association of Universities for Research in Astronomy and the NSF.

This work was funded by the Spanish grant AYA2004-04462. SGK was supported by
NASA grants NAG5--13501 \& NNG05GD58G and by NSF grant ATM--0318390.

\begin{table*}
  \begin{center}
    \caption{Description of the data sets used in this work.
}\vspace{1em}
    \renewcommand{\arraystretch}{1.2}
    \begin{tabular}[h]{lcclc} \hline
  Instrument & Degree      & Frequency (mHz)   &  Reference                  & Referred as \\ \hline
  MDI    & $4\le\ell\le25$ & $1.1\le\nu\le3.9$ & Korzennik 2005              & {\sf KM} \\
  GONG   & $4\le\ell\le25$ & $1.1\le\nu\le3.9$ & Korzennik 2005              & {\sf KG} \\
  MDI    & $5\le\ell\le25$ & $1.7\le\nu\le3.9$ & Schou \etal\ 1998           & {\sf SM} \\
  MDI    & $1\le\ell\le10$ & $1.8\le\nu\le3.9$ & Jim\'enez-Reyes \etal\ 2000 & {\sf JM} \\
  GOLF   & $1\le\ell\le 3$ & $1.1\le\nu\le2.2$ & Garc\'\i a \etal\ 2004      & {\sf GG} \\
  GOLF   & $\ell=1$        & $1.4\le\nu\le3.9$ & Gelly \etal\ 2002           & {\sf BG} \\
      \hline \\
      \end{tabular}
    \label{tab:table1}
  \end{center}
\end{table*}

\begin{table*}
  \begin{center}
    \caption{Description of the combined data set used,
 and a brief description of the combination procedure.}\vspace{1em}
    \renewcommand{\arraystretch}{1.2}
    \begin{tabular}[h]{lccl}\hline 
  Instrument & Degree          & Frequency  (mHz)  & Combination Procedure \\\hline
  GOLF      &  $\ell=1$        & $1.1\le\nu\le2.2$ & Average of {\sf GG} and {\sf BG}  \\
  GOLF      &  $\ell=2,3$        & $1.1\le\nu\le2.2$ & Data from {\sf GG}  \\
  MDI      &  $\ell=1,3$        & $2.2 <\nu\le3.9$ & Data from {\sf JM}  \\
  MDI-GONG  & $ 4\le\ell\le10$ & $1.1\le\nu\le1.8$ & Average of {\sf KM} and {\sf KG}  \\
  MDI-GONG  & $ 4\le\ell\le10$ & $1.8\le\nu\le3.9$ & Average of {\sf JM, KM, SM} and {\sf KG}  \\
  MDI-GONG  & $11\le\ell\le25$ & $1.1\le\nu\le1.8$ & Average of {\sf KM} and {\sf KG}  \\
  MDI-GONG  & $11\le\ell\le25$ & $1.8\le\nu\le3.9$ & Average of {\sf KM, KG} and {\sf SM}  \\\hline \\
      \end{tabular}
    \label{tab:table2}

{\small
The MDI-GONG-GOLF set consists of the ensemble of these values, although the MDI data for $\ell=1,3$ 
and $2.2 <\nu\le3.9$ are only used to compute artificial data sets.}
  \end{center}
\end{table*}

\begin{table*}
  \begin{center} 
\caption{Description of the artificial data sets used to study the sensitivity
of helioseismic data to the dynamics of the solar core.}
\vspace{1em}
\renewcommand{\arraystretch}{1.2}
\begin{tabular}[h]{lcccc}

\hline ~        &  ~   & \multicolumn{3}{c}{Freq.~range (mHz)} \\
\hline Data set & Note & $\ell=1$         &  $\ell = 2, 3$  &  $\ell > 3$ \\\hline
 Set {\Aone} &  ~   & $1\le\nu\le2.2  $  & $1\le\nu\le2.2  $ & $1\le\nu\le3.9$  \\
 Set {\Atwo} &  ~   & $1\le\nu\le3.9$  & $1\le\nu\le3.9$ & $1\le\nu\le3.9$  \\
 Set {\Athree} & (1)  & $1\le\nu\le3.9$  & $1\le\nu\le3.9$ & $1\le\nu\le3.9$  \\
 Set {\Afour} & (2)  & $0.1\le\nu\le2.2$  & $0.1\le\nu\le2.2$ & $1\le\nu\le3.9$ \\\hline \\
      \end{tabular}
    \label{tab:table3}

{\small
$^{(1)}$ Data set {\Athree} differs from data set {\Atwo} in its
uncertainties: the uncertainties for $\ell < 4$ are set to be the ones for the
sectoral splittings of $\ell=25$.

$^{(2)}$ One $\ell=1$ and one sectoral $\ell=2$ $g$-modes rotational
splittings were added to the {\Aone} data set.
}
  \end{center}
\end{table*}

\begin{figure}[tbp*]
\begin{center}
\includegraphics[width=15pc,angle =90]{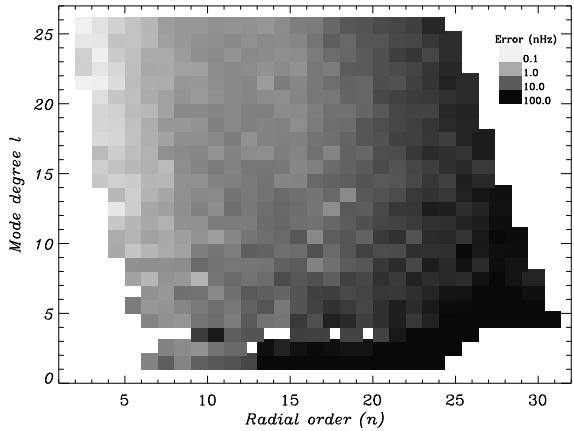}
\caption{Observational uncertainties of the frequency splittings, shown as a
function of radial order $n$ and spherical harmonic degree, $\ell$.
Uncertainties obtained from 2088-days long time-series for the combined
sectoral frequency splittings set are shown using a logarithmic gray scale.}
\end{center}
\end{figure} 

\begin{figure}[tbp*]
\begin{center}
\includegraphics[width=15pc,angle =90]{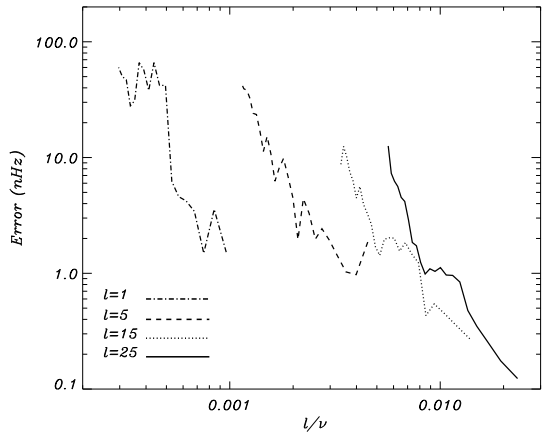}
\caption{Observational uncertainties of sectoral frequency splittings as a
function of the inner turning point proxy $\ell/\nu$, for a selected set of
degrees. The solid, dotted, dashed and dashed-dotted lines correspond to
$\ell=25,15,5$ and $1$, respectively.}
\end{center}
\end{figure}

\begin{figure}[tbp*]
\begin{center}
\includegraphics[width=15pc,angle =90]{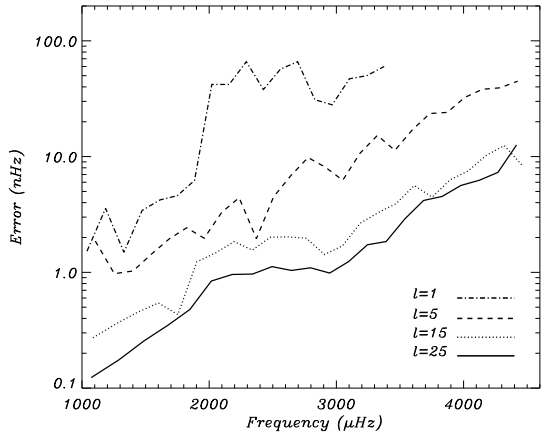}
\caption{Observational uncertainties of sectoral frequency splittings as a
function of the mode frequency, $\nu$, for a selected set of degrees. The
solid, dotted, dashed and dashed-dotted lines correspond to $\ell=25,15,5$ and
$1$, respectively.}
\end{center}
\end{figure}

\begin{figure}[tbp*]
\begin{center}
\includegraphics[width=15pc,angle =0]{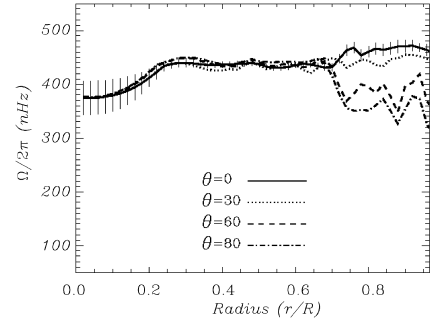}
\caption{Inversion of the combined MDI-GOLF-GONG data set (see Table~2),
plotted as a function of radius at several latitudes, $\theta$. The solid,
dotted, dashed and dotted-dashed lines correspond to the rotation rate for the
equator, and at latitudes of $30^o$, $60^o$ and $80^o$, respectively. Error
bars are shown for the rotation rate at the equator, whereas the error
distributions for other latitudes are shown in Fig.~12.}
\end{center}
\end{figure}
 
\begin{figure}[tbp*]
\begin{center}
\includegraphics[width=15pc,angle =0]{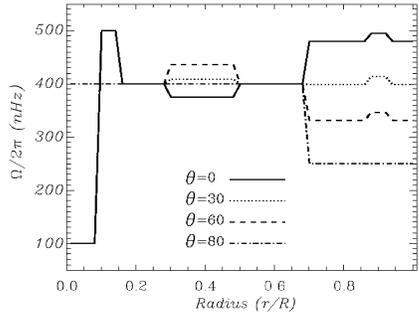}
\caption{Artificial rotation rate for the solar interior shown as a function
of radius, $r$, at different latitudes, $\theta$. This artificial profile
incorporates latitudinal variations in two zones to test the inversion
sensitivity to latitude.  The solid, dotted, dashed and dotted-dashed
correspond to the rotation rate for the equator, and at latitudes of $30^o$,
$60^o$ and $80^o$, respectively.}
\end{center}
\end{figure}

\begin{figure}[tbp*]
\begin{center}
\includegraphics[width=15pc,angle =0]{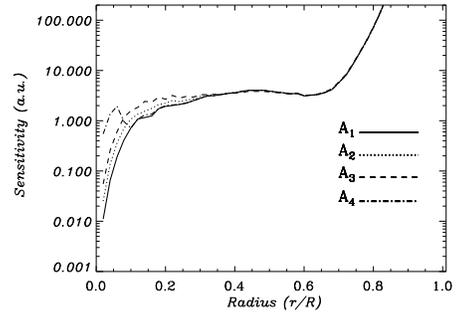}
\caption{Sensitivities of the solar rotation at the equator for sets \Aone\ to
\Afour, represented by solid lines, dotted lines, dashed lines, and
dotted-dashed lines, respectively.}
\end{center}
\end{figure}

\begin{figure}[tbp*]
\begin{center}
\includegraphics[width=15pc,angle =0]{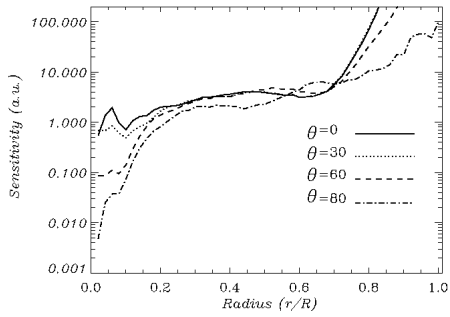}
\caption{Sensitivities to set \Afour\ of the solar rotation at different
latitudes, namely the equator (solid line), $30^o$ (dotted lines), $60^o$
(dashed lines) and $80^o$ (dotted-dashed lines). }
\end{center}
\end{figure}

\begin{figure}[tbp*]
\begin{center}
\includegraphics[width=15pc,angle =0]{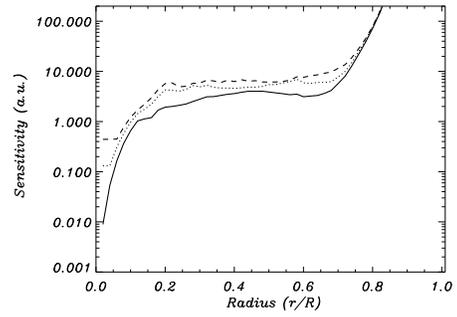}
\caption{Sensitivities to set \Aone\ of the solar rotation at the equator for 
three different smoothing matrices $H$, namely those corresponding to 
 first differences (solid line), second differences (dotted lines) 
and third differences (dashed lines).}
\end{center}
\end{figure}

\begin{figure}[tbp*]
\begin{center}
\includegraphics[width=15pc,angle =0]{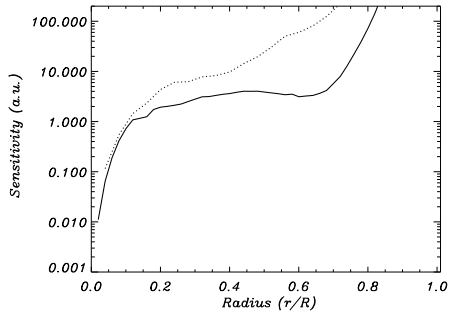}
\caption{Sensitivities to set \Aone\ of the solar rotation at the equator for
two different distributions of model grid points, namely 50 \& 16 points
(solid line) and 25 \& 16 points in the radial and angular directions
respectively (dotted line).}
\end{center}
\end{figure}

\begin{figure}[tbp*]
\begin{center}
\includegraphics[width=15pc,angle =0]{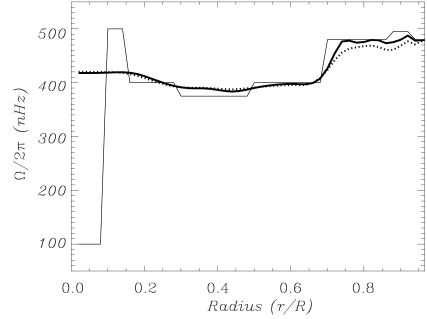}
\caption{Inversions of set \Athree\ as a function of radius at the
equator. Results for the standard RLS inversion (no vector $W$) and for the
newly developed inversion method are shown in dotted and solid lines,
respectively.  The artificial rotation rate used to calculate the input sets
is shown as the thin solid line. }
\end{center}
\end{figure}

\begin{figure}[tbp*]
\begin{center}
\includegraphics[width=15pc,angle =0]{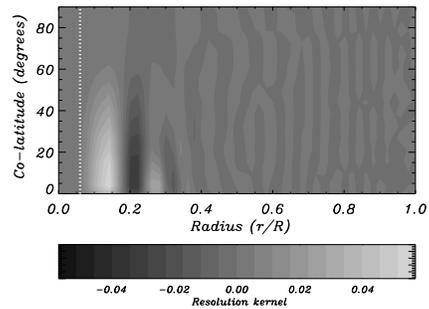}
\caption{Plot of the resolution vector $R_{ij}$, $j=1,N$, calculated for set
\Aone\ corresponding to the inversion estimated at the equator for $r_i=0.06
\RSun$ (indicated by vertical dotted lines). The same result was obtained for the inversion of \Atwo\ .}
\end{center}
\end{figure}

\begin{figure}[tbp*]
\begin{center}
\includegraphics[width=15pc,angle =0]{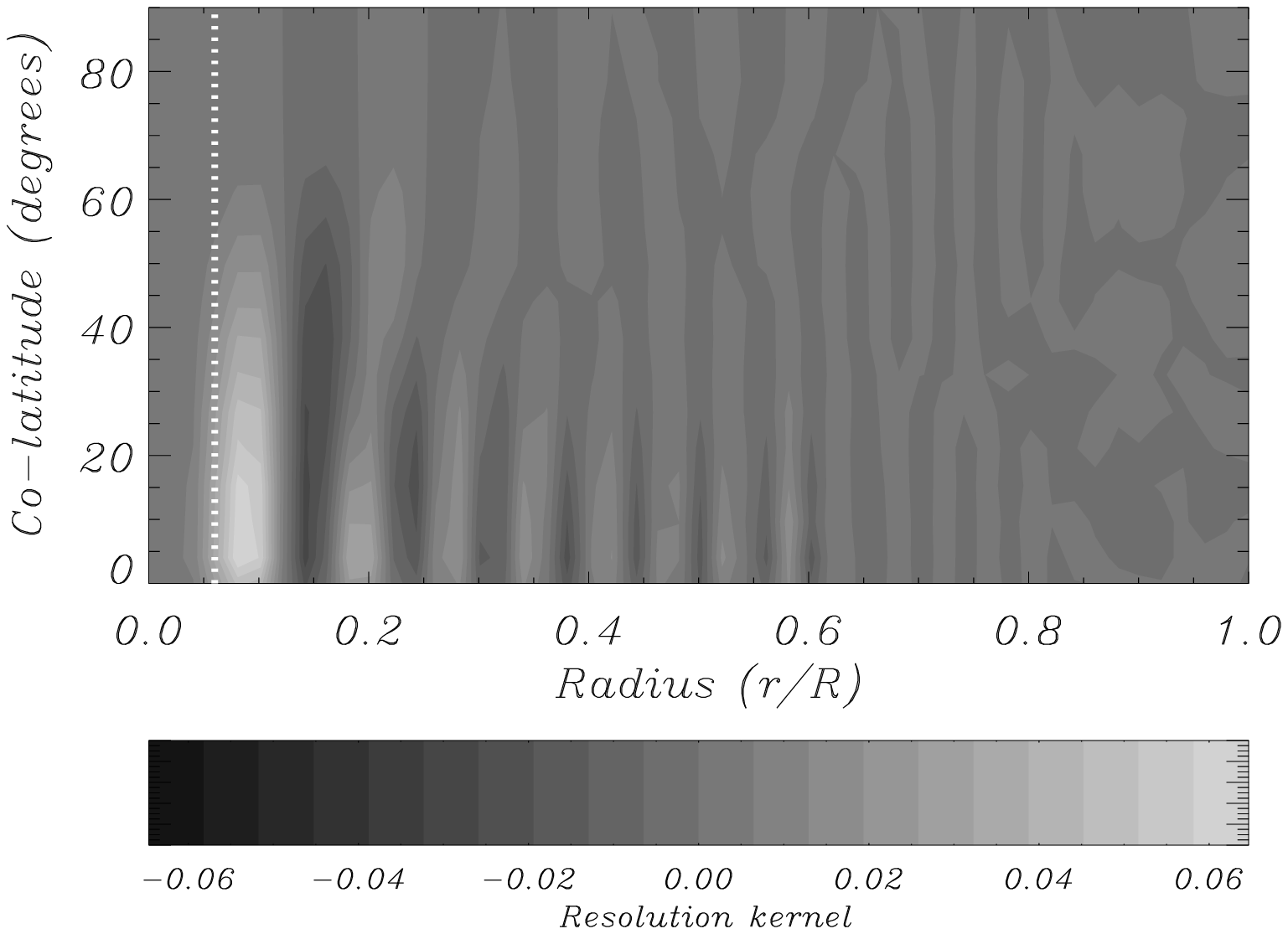}
\caption{Plot of the resolution vector $R_{ij}$, $j=1,N$, calculated for set
\Athree\ corresponding to the inversion estimated, as in Fig.~11, at the
equator for $r_i=0.06 \RSun$.}
\end{center}
\end{figure}

\begin{figure}[tbp*]
\begin{center}
\includegraphics[width=15pc,angle =0]{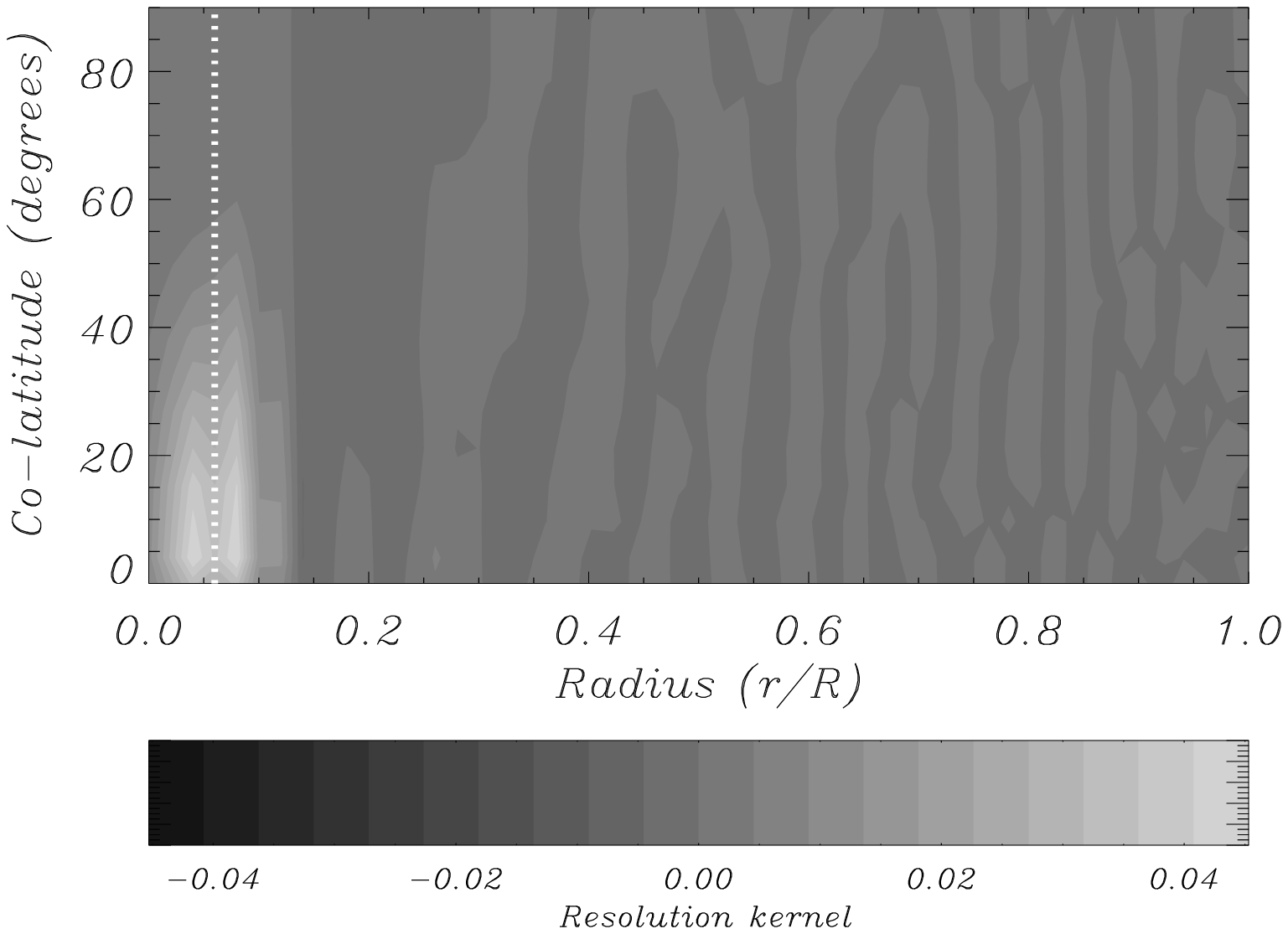}
\caption{Plot of the resolution vector $R_{ij}$, $j=1,N$, calculated for set
\Afour\ corresponding to the inversion estimated , as in Figs.~11 \& 12, at
the equator for $r_i=0.06 \RSun$.}
\end{center}
\end{figure} 

\begin{figure}[tbp*]
\begin{center}
\includegraphics[width=15pc,angle =0]{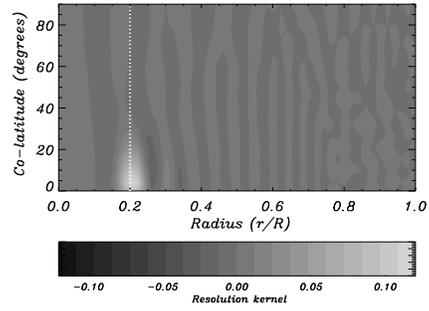}
\caption{Plot of the resolution vector $R_{ij}$, $j=1,N$, calculated for set
\Athree\ corresponding to the inversion estimated at the equator but for
$r_i=0.20 \RSun$ (indicated by vertical dotted lines).}
\end{center}
\end{figure} 

\clearpage

\begin{figure}[tbp*]
\begin{center}
\includegraphics[width=15pc,angle =0]{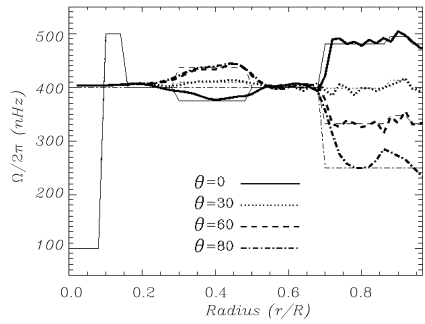}
\caption{Inversion of set \Aone\ as a function of radius at different
latitudes, namely the equator (solid line), $30^o$ (dotted lines), $60^o$
(dashed lines) and $80^o$ (dotted-dashed lines). The artificial rotation rate
used to calculate the input set is shown as thin lines. }
\end{center}
\end{figure} 

\begin{figure}[tbp*]
\begin{center}
\includegraphics[width=15pc,angle =0]{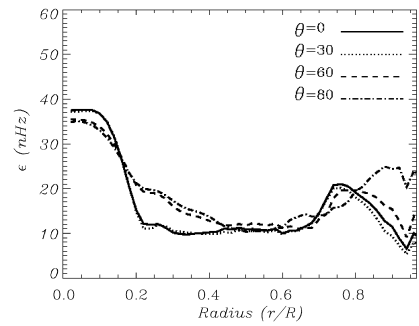}
\caption{Error distribution for the inversion of set \Aone\ as a function of
radius at different latitudes, namely the equator (solid line), $30^o$ (dotted
lines), $60^o$ (dashed lines) and $80^o$ (dotted-dashed lines).}
\end{center}
\end{figure}

\begin{figure}[tbp*]
\begin{center}
\includegraphics[width=15pc,angle =0]{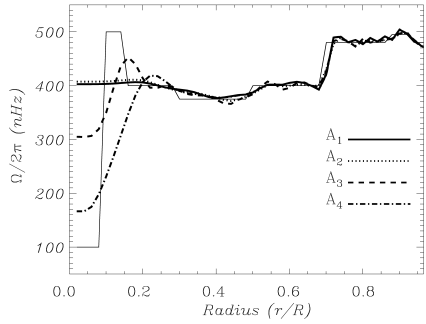}
\caption{Inversions of set \Aone\ to \Afour\ plotted as a function of radius
at the equator. Results for sets \Aone\ to \Afour\ are shown as solid, dotted,
dashed and dotted-dashed lines respectively. The artificial rotation rate used
to calculate the input sets is shown as a thin solid line.}
\end{center}
\end{figure}

\begin{figure}[tbp*]
\begin{center}
\includegraphics[width=15pc,angle =0]{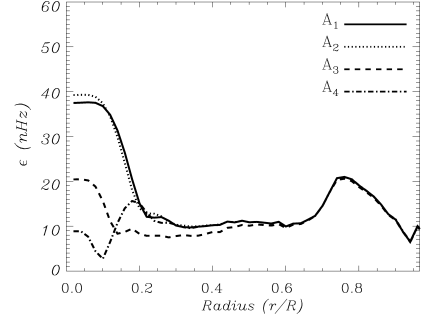}
\caption{Error distribution for the inversions of set \Aone\ to \Afour\
plotted as a function of radius at the equator. Results for sets \Aone\ to
\Afour\ are shown as solid, dotted, dashed and dotted-dashed lines
respectively. }
\end{center}
\end{figure}

\begin{figure}[tbp*]
\begin{center}
\includegraphics[width=15pc,angle =90]{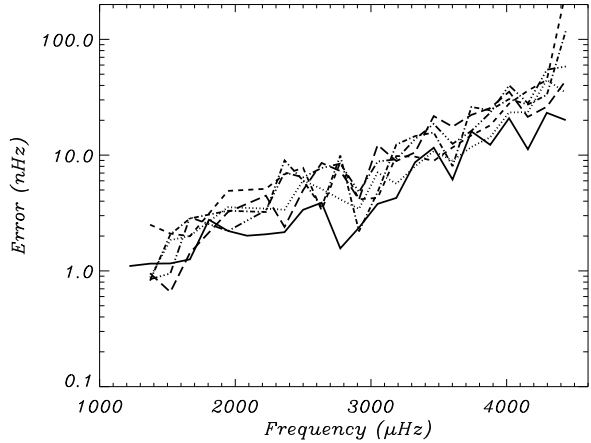}
\caption{Observational uncertainties for $\ell=25$ sectoral frequency
splittings as a function of the mode frequency, $\nu$. The solid line
represents the uncertainties for the 2088-days long data set, whereas the
other line styles represent the uncertainties for the five 728-days long data
sets. }
\end{center}
\end{figure}

\end{document}